# Hidden symmetries in primary sequences of small α proteins

Ruizhen Xu, Yanzhao Huang, Mingfen Li, Hanlin Chen, and Yi Xiao[*]

*Biomolecular Physics and Modeling Group, Department of Physics, Huazhong University of Science and Technology, Wuhan 430074, Hubei, China*

**Abstract**

Proteins have regular tertiary structures but irregular amino acid sequences. This made it very difficult to decode the structural information in the protein sequences. Here we demonstrate that many small α protein domains have hidden sequence symmetries characteristic of their pseudo-symmetric tertiary structures. We also present a modified method of recurrent plot to reveal this kind of the hidden sequence symmetry. The results may enable us understand parts of the relations between protein sequences and their tertiary structures, i.e, how the primary sequence of a protein determines its tertiary structure.

**Introduction**

One of unsolved problems in molecular biophysics is how proteins encode their structural information in their amino acid sequences [1-7]. The amino acid sequences of proteins appear very irregular, but the three-dimensional structures they encode clearly show certain regularity. For an example, many proteins have symmetric tertiary structures but a seemly random primary sequences, i.e., amino acid sequences. This riddle has motivated intensive studies of the correlation properties of protein sequences [8–15] to see whether they are random or not. However, these studies gave opposing results: some studies showed that protein sequences were indistinguishable from random ones, while other results indicated that protein sequences were nonrandom. For examples, White and Jacobs [8][9] studied the statistical distribution of hydrophobic residues along the length of protein chains by using a binary hydrophobicity scale, which assigns hydrophobic residues a value of one and nonhydrophobes a value of zero. Using the standard run test, they found that, for the majority of the 5247 proteins examined, the distribution of hydrophobic residues along a sequence could not be distinguished from that expected for a random distribution. On the other hand, Pande et al. [12] studied the statistics of protein sequences by using the idea of mapping the sequence onto the trajectory of a random walk. They found pronounced deviations from pure randomness. It is noted that both studies use a binary scale of hydrophobicity and hydrophilicity but different mapping schemes. In the work of White and Jocobs, Phe, Met, Leu, Ile, Val, Cys, Ala, Pro, Gly, Trp and Tyr were considered as hydrophobic and other residues as hydrophilic, while

---

in the work of Pande et al., Lys, Arg, His, Asp and Glu were considered as hydrophilic and other as hydrophobic. Weiss and Herzel [14] analyzed the correlation functions in large sets of nonhomologous protein sequences. They found that the hydrophobicity autocorrelation showed period 3 to 4 oscillations. These oscillations decayed until they vanish at a length of 10–15 amino acids and they can be related to the 3.6 periodicity of α-helices.

In 1998, Rackovsky[15] tried to solve the contradiction above and demonstrated the existence in protein domain sequences of "hidden" periodic signals, characteristic of the architectures of those domains. The characteristic signals define sequence units, which may correspond to specific structural features. His study on two different architectures (the TIM barrel and neuraminidase folds) suggested that symmetric structures are characterized by sets of sequence signals, which are members of harmonic series and which therefore impose commensurate periodicities on the structure. Asymmetric structures are characterized by mutually incommensurate sets of sequence signals. Rackovsky's results imply that the protein sequences may have "hidden" regular patterns characteristic of their tertiary structures, although these definite patterns characteristic of a particular architecture can occur in different physical property for different protein sequences.

In the present work, we show that the primary sequences of many small α protein domains have a kind of hidden symmetry characteristic of their tertiary structures even without considering their physical properties and we also present a modified method of recurrence plot to demonstrate this kind of hidden symmetry.

## Methods

Symmetry means similarity. For an example, the sequence, ADJGFADJGFADJGF, is composed of three identical subsequences and we say that it has exact three-fold symmetry. But real protein sequences do not have such exact symmetry. In fact, protein sequences appear nearly random as indicated above. However, to make structures symmetric does not need the sequences to have exact symmetry, i.e., to be composed of identical parts. It is known from sequence alignment [16-17] that if two proteins have more than 25% identical amino acids, they likely have similar tertiary structures [18][19]. We use this rule to define the symmetry of the primary sequences, i.e., a protein sequence is said to have N-fold pseudo-symmetry if it is composed of N subsequences with certain number of (usually more than 20%) identical amino acids.

The hidden symmetries of the primary sequences can be shown by using a modified recurrence quantification analysis. Recurrence quantification analysis is a relatively new nonlinear technique [20-22], but its original form is not convenient to show how the recurrence changes with the lengths of segments. In order to show the symmetric patterns in the protein sequences, we developed a modified version of recurrence plot, i.e., a kind of subsequence alignment in a primary sequence as follows.

For an arbitrary protein sequence $S = x_1 x_2 x_3 ... x_N$, one constructs a set of $d$

-dimensional vectors (embedding space):

$$X_1 = (x_1, x_2, ..., x_d),$$
$$X_2 = (x_2, x_3, ..., x_{d+1}),$$
$$......$$
$$X_{N-d+1} = (x_{N-d+1}, x_{N-d+2}, ..., x_N) \quad (1)$$

which corresponds to all possible segments of $d$ consecutive symbols. The modified recurrence plot is built as follows. The horizontal axis of the modified recurrence plot is the residue index in the sequence and the vertical axis is the embedding dimension $d$ (the length of the segments). A point is placed at $(i, d)$ if $X_i$ and $X_j$ are similar.

A segment $X_j$ is similar to $X_i$ if $D(X_i, X_j) \geq r$ where $D(X_i, X_j)$ represents the average distance between $X_i$ and $X_j$ which is defined here by the average Hamming distance:

$$D(X_i, X_j) = \sum_{k=0}^{d-1} h(x_{i+k}, x_{j+k})/d \quad (2)$$

where

$$h(x_i, x_j) = \begin{cases} 1 & x_i = x_j \\ 0 & x_i \neq x_j \end{cases} \quad (3)$$

The degree of the similarity of the segments is determined by $r$ ($0 \leq r \leq 1$). For examples, two segments are required to be identical if setting $r = 1$ and to have at least 25% identical amino acids if setting $r = 0.25$.

The usefulness of this method can be demonstrated by applying to an artificial periodic sequence, <u>ACDEFGHIKLMNPQRSTVWY</u><u>ACDEFGHIKLMNPQRSTVWY</u><u>ACDEFGHIKLMNPQRSTVWY</u>. The sequence has three identical subsequences and each of them consists of 20 elements. Fig.1 is the modified recurrence plot of this three-fold symmetric sequence. It shows that the modified recurrence plot can not only find the locations of similar subsequences but can also give their lengths. The plot is made of three identical right-angled triangles. This indicates that the three segments (from 1 to 20, 21 to 40, and 41 to 60) of length 20 are identical with each other. In fact, the perfect right-angled triangles imply that the subsequences of any lengths in these three segments are also identical. By the way, since this is an ideal periodic sequence, the results don't depend on the choice of the similarity $r$.

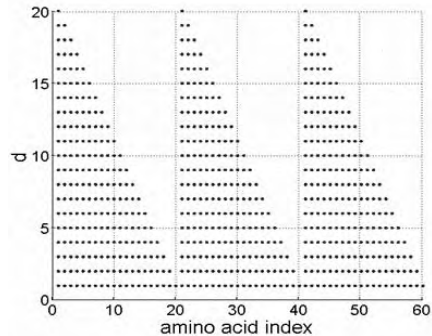

Fig.1 The modified recurrence plot of an artificial sequence with three-periodicity. The plot is divided into three identical parts and shows an ideal three-fold symmetry.

**Results and Discussions**

For native proteins, their amino acid sequences do not show the exact symmetries as above and in fact they appear nearly "random" in usual sense. However, it is a too strong definition for the sequence similarity that all of their amino acids must be identical, i.e., $r=1$. As indicated above, if two proteins have more than 25% identical amino acids, they likely have similar tertiary structures. So a suitable definition of the sequence similarity may reveal the hidden sequence symmetries characteristic of the tertiary structures. In the present work we investigate the hidden symmetry of the typical sequences of three types of folds: up-down bundle, orthogonal bundle and irregular bundle. The basic architectures of these folds have a common topological structure, i.e., they all have a hairpin-like structure but with two arms being arranged in different orientations (Fig.2-Fig.15). The two arms usually have similar secondary structures, i.e., two α helices with similar lengths, although sometimes they may be a little different. However, from the point of view of topology, the structures of the two arms are similar or equivalent and so the tertiary structures of these folds can be considered to have two-fold symmetry.

A typical protein domain of these folds is transferase (PDB id: 1r2a) with two α helices with similar length and it belongs to the up-down bundle fold (Fig.2a). The tertiary structure of 1r2a is very regular and can be regarded to be made of two similar substructures, i.e., it can be considered to have a pseudo two-fold symmetry. However, its primary sequence appears irregular (Fig.2b). This can be seen from the modified recurrence plot for larger $r$ (Fig.2c). Fig. 2c shows clearly that the sequence of 1r2a does not reveal any symmetry if $r=0.70$, which means that the similar subsequences must have at least 70% identical amino acids. How does this irregular primary sequence determine a symmetric tertiary structure? In fact, the symmetry of the primary sequence can be exhibited when the similarity $r$ is given a suitable value. Fig.2c shows that a two-fold symmetry emerges in the primary sequence as r decreases and the sequence is divided into two similar subsequences when $r$ comes to 0.25. When $r$ is set to be 0.25, two subsequences are defined to be similar as long as they have more than 25% identical amino acids.

Fig.3 shows the more detailed information about the similar subsequences of 1r2a for $r=0.25$. Fig.3c shows the length and the start and end positions of these two similar segments. In the plot there are two peaks located at (1, 20) and (20, 20) and, as mentioned above, it means there are two similar subsequences of length 20 and beginning at 1st and 20th amino acids respectively. Fig.3b shows the primary and secondary structures corresponding to these two similar subsequences. The upper figure shows the primary and secondary structures of the first subsequence from 1st to 19th amino acids and the lower figure shows those of the second one from 20th to 40th. It is clear that the two subsequences are connected at the loop region and they have a very good correspondence to the two similar substructures, i.e., the two α helices with similar length. This means that the primary sequence of 1r2a shows the same symmetry as its tertiary structure. This kind of symmetry has not been found by other methods. Furthermore, Fig. 3a shows the names and the positions of five identical amino acids of the two subsequences in the tertiary structure of 1r2a. The five identical amino acids are Q (Gln), P (Pro), P (Pro), L(Leu) and Y(Tyr). It is found that four of the identical amino acids are located at the loop region and the end regions of the α helices and one of them at the middle of the α helices. What is the role of these identical amino acids? Are they the key amino acids of the formation of the secondary or tertiary structures or those for the stability of the tertiary structure or for the functions? These problems are worthy to be investigated in the future.

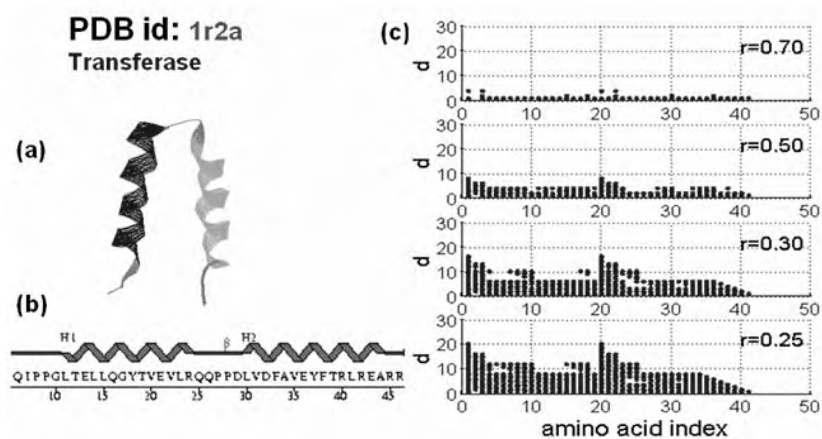

Fig.2 The structures and recurrence plots of transferase (1r2a): (a) The tertiary structure with two-fold symmetry; (b) The primary and secondary structures(taken from PDBsum); (c) The modified recurrence plots with different values of $r$. It shows that the symmetry of the primary sequence emerges as the value of $r$ decreases.

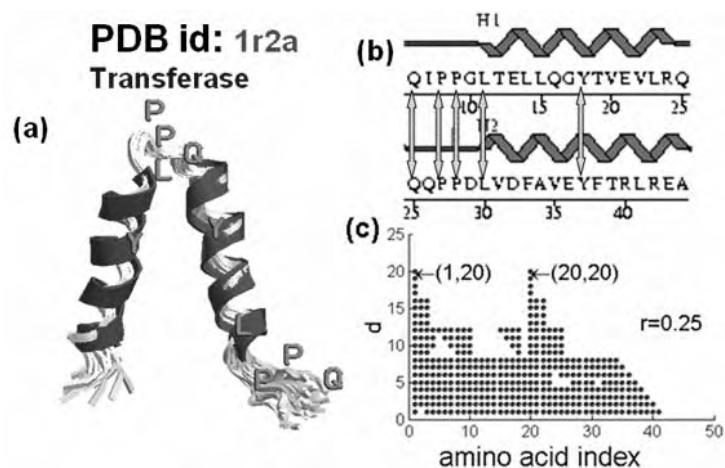

Fig.3 The structures and the modified recurrence plot of transferase (1r2a): (a)

The tertiary structure with the labeled names and positions of the identical amino acids; (b) The primary and secondary structures of the two similar subsequences and the identical amino acids are showed by arrows; (c) The modified recurrence plot when $r$=0.25 and in the plot the locations and the length of two similar subsequences are indicated.

Fig.4 and Fig.5 show the structures and the modified recurrence plots of other non-homologous sequences of the protein domains (PDB ID: 1m1e, 1koy and 1dav) and of the up-down fold. As in Fig.3, these figures show the tertiary structures, the primary and secondary structures of the similar subsequences, and modified recurrence plots. The values of similarity $r$ used in each calculation are also shown in the modified recurrence plot respectively.

Similarly, Fig.6 to Fig.8 and Fig.9 to Fig.13 show the structures and the modified recurrence plots of the typical non-homologous sequences of the orthogonal bundle fold (PDB ID: 1irq, 2cpg and 1mdy) and the irregular up-down fold (PDB ID: 1nkp, 1amg, 1ebd, and 2pdd), respectively.

All of the protein domains shown in Fig.4 to Fig.13 reveal two-fold pseudo symmetries in their sequences, i.e., they are made of two similar subsequences. As for 1r2a, it is found that, for all the protein domains considered here, the two similar subsequences are connected at the loop regions between the two α helices. This means that each of the two similar subsequences corresponds to one of the similar substructures, i.e., the subsequence and the substructure have a good correspondence. As pointed out above, the two similar substructures may not have the exact secondary structures, although they usually have in the most cases. They have similar topological structures. For an example, the protein domain 1irq (or 2cpg) (Fig.7) has one helix in one of the arms of the hairpin-like structure and another arm of the hairpin has both a strand and a helix. However, in the topological sense, the two arms are similar and can be considered to be symmetric. This is in agreement with Baker's opinion that the primary sequence of a protein only determines its topological structure [23]. It is also noted that there are very short α helices in some protein domains (e.g., 1dav, 1ebd and so on) and they can be ignored or considered as one part of a long α helix.

In the figures, we also indicated the names and locations of the identical amino acids of the two similar subsequences. As in the Fig.3, most of the identical amino acids are located at the loop regions or the end regions of the α helices. Furthermore, it is found that the main types of the identical amino acids are Lys (K), Arg (R) and Leu (L).

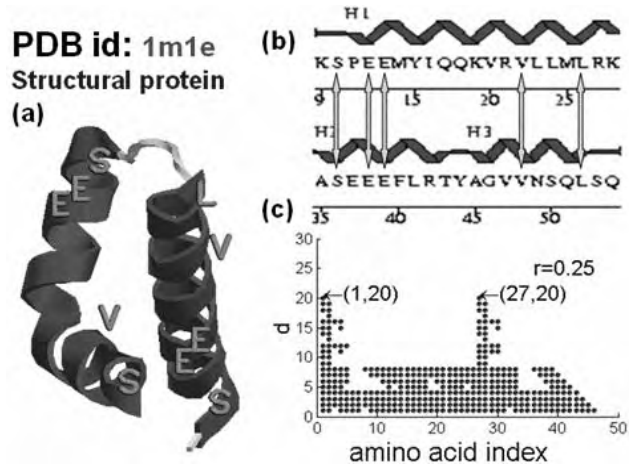

Fig.4 The same as Fig.3 but for a structural protein (1m1e).

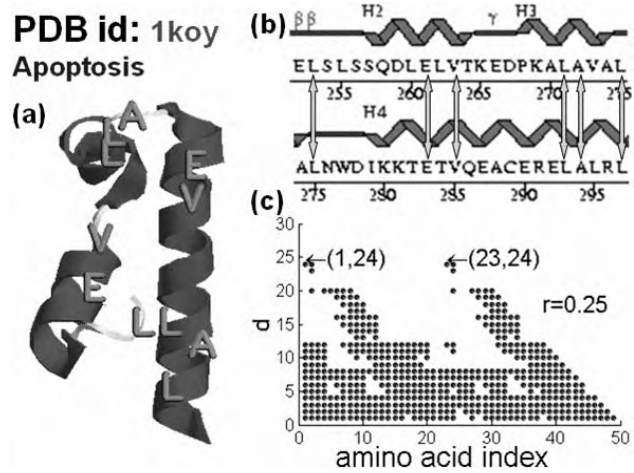

Fig.5 The same as Fig.3 but for apoptosis (1koy).

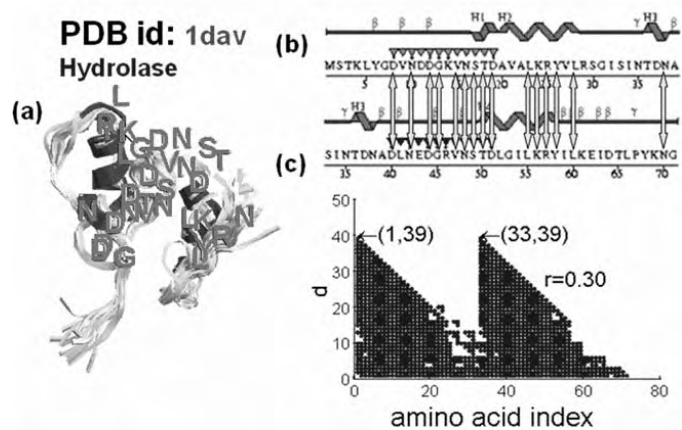

Fig.6 The same as Fig.3 but for hydrolase (1dav) and $r$=0.30.

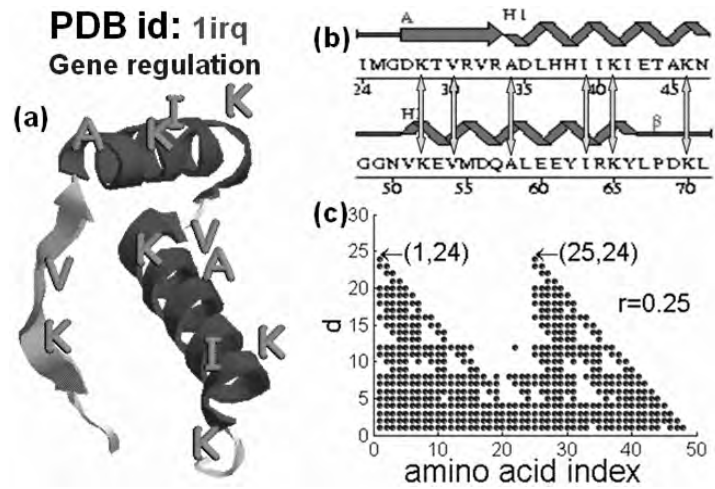

Fig.7 The same as Fig.3 but for gene regulation (1irq).

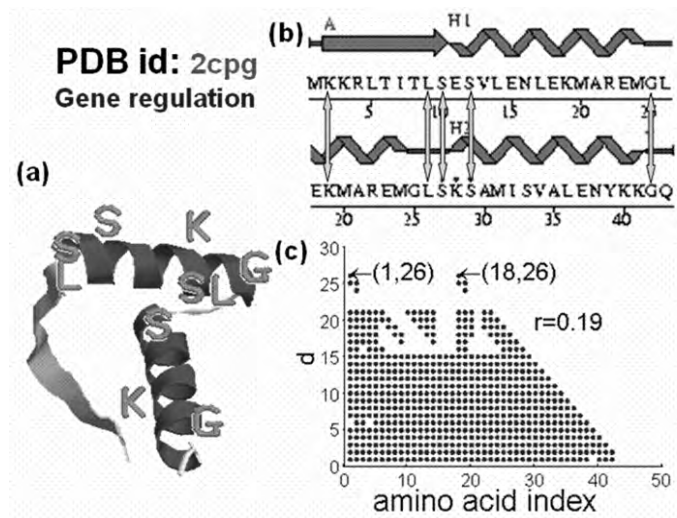

Fig.8 The same as Fig.3 but for gene regulation (2cpg) and $r$=0.19.

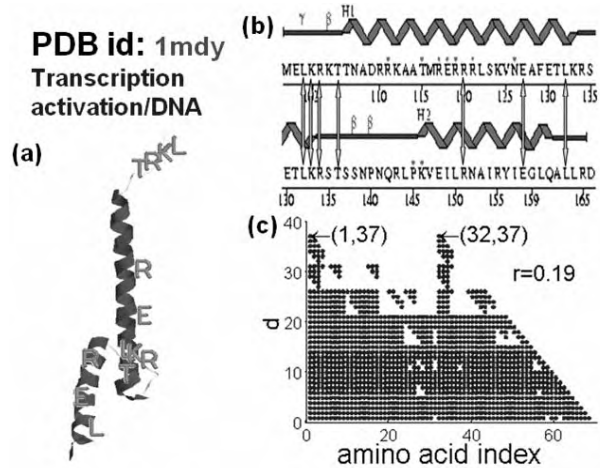

Fig.9 The same as Fig.3 but for transcription activation/DNA (1mdy) and $r$=0.19.

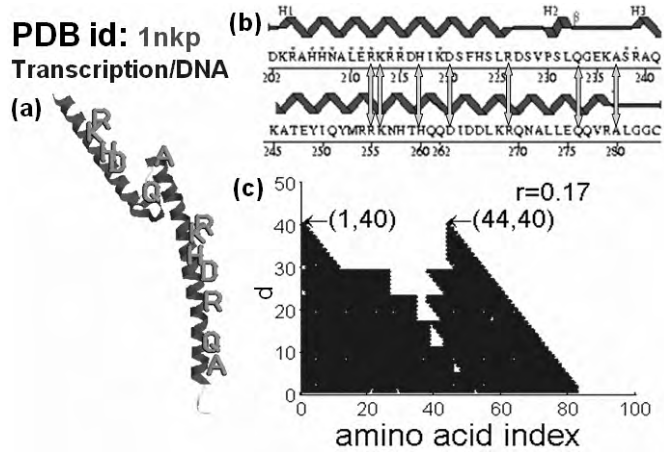

Fig.10 The same as Fig.3 but for transcription/DNA (1nkp) and *r*=0.17.

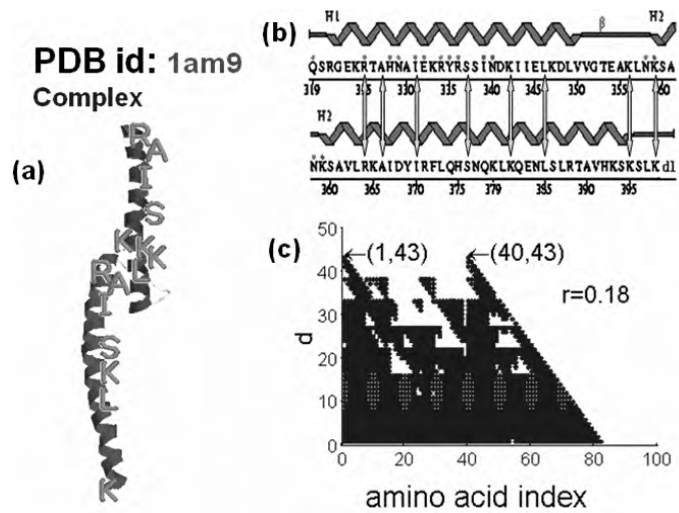

Fig.11 The same as Fig.3 but for complex (1am9) and *r*=0.18.

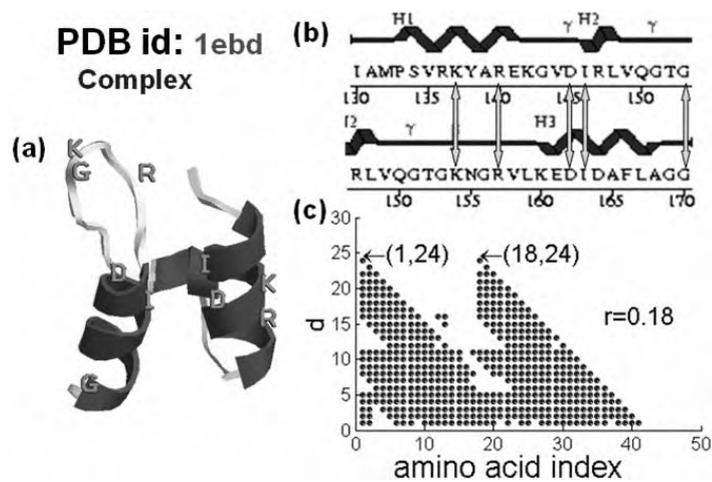

Fig.12 The same as Fig.3 but for complex (1ebd) and *r*=0.18.

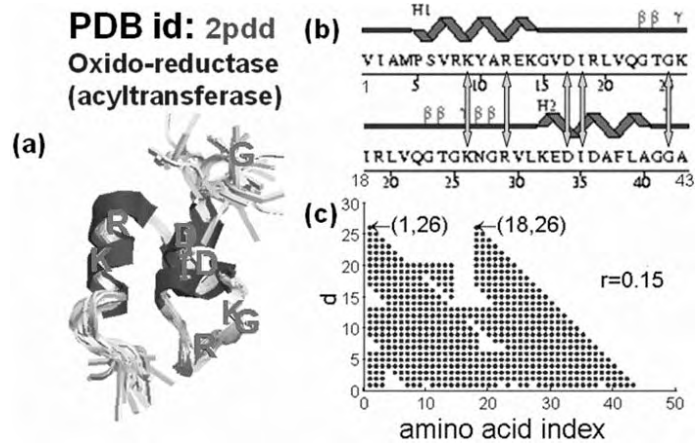

Fig.13 The same as Fig.3 but for oxido-reductase (2pdd) and *r*=0.15.

In the present paper, we only use the simplest definition of the sequence similarity, i.e., by comparing the number of identical amino acids in two subsequences. Using this definition, we didn't find the primary sequence symmetries of some folds with symmetric structures, e.g., 1k4t (Fig.14). This may be due to that the symbolic sequence alignment used here didn't take account of the similarity between the amino acids in physicochemical properties, e.g., hydrophobicity, volumes, frequencies of occurrence in α-helix and β strand and so on. As shown by Rackovsky [15], the tertiary structures of proteins, even those with the same architecture, may not be determined by one factor. So, a detailed analysis of the primary sequence symmetry based on various physiochemical properties will be carried out in the future.

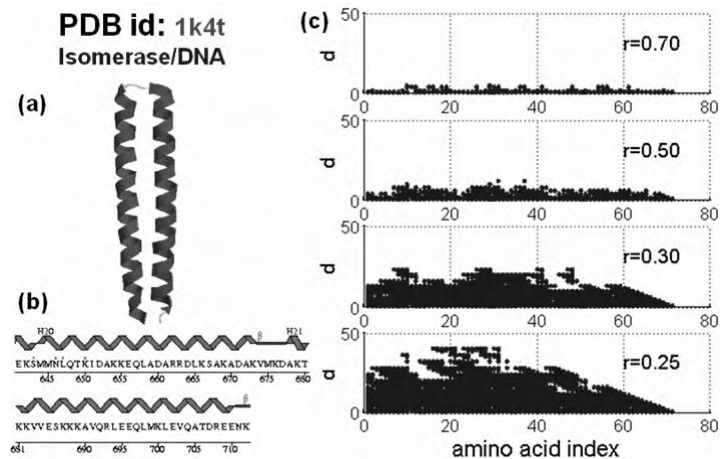

Fig.14 The structures and recurrence plots of Isomerase/DNA (1k4t): (a) The tertiary structure with two-fold symmetry; (b) The primary and secondary structures; (c) The modified recurrence plots with different values of *r*. It can not reveal the symmetry of the primary sequence as the value of *r* decreases.

The results above indicate that all of the protein sequences considered here have two-fold hidden symmetries. Furthermore, these symmetries are the same as their tertiary structures. These protein sequences are typical ones and represent most of the protein sequences of the architectures we considered in the present paper. So it may suggest that the formations of the symmetric tertiary structures of these protein domains are the results of their sequence symmetries, i.e., the symmetries of these tertiary structures are encoded by the sequences. This may be one of the manifests of the generally-accepted dogma that the primary structure of a protein determines its tertiary structure. This result may have some implications: (1) The local structures of proteins may be determined by the sequences using the same role as the global structures, i.e., as long as two sequences have certain number of identical amino acids (usually 25%), they are likely to have similar tertiary structures. This satisfies the Nature's economy principle. (2) It is known that different sequences may have the similar symmetric tertiary structure. One of the reasons may be that they have common hidden sequence symmetry. (3) The sequences of the proteins with symmetric tertiary structures are made of similar subsequences and may be formed by duplication. However, since the identical amino acids may be only 25%, i.e., the similar subsequences only have a small part of amino acids which are identical, this can also suggest that the formation of similar subsequences in the protein sequences may be a result of the point mutations. (4) If the substructures of the protein domains are mainly determined by the identical amino acids, they may play important roles in the domain folding, stability or functions.

**Acknowledgement:** This work is supported by the National Natural Science Foundation of China under Grant No. 90103031 and the Foundation of the Ministry of Education.